\def\final{0}

\documentclass[superscriptaddress, 
notitlepage, amsmath, amssymb, aps,
pra,
 10pt, letterpaper]{revtex4-1}
\usepackage{graphicx}
\usepackage{dcolumn}
\usepackage{bm}

\usepackage{fancyhdr}
\usepackage{latexsym}
\usepackage{graphicx,color}
\usepackage{epstopdf}  
\usepackage[pdfstartview=FitH]{hyperref}
\usepackage{xargs}
\usepackage{shadow,epsf,amsthm,amssymb,amsmath}

\usepackage{enumerate}

\usepackage{setspace}                           

\usepackage{lipsum}

\usepackage{tikz}	
\usetikzlibrary{backgrounds,fit,decorations.pathreplacing}  

\usepackage[colorinlistoftodos,prependcaption]{todonotes}

\newtheorem{definitionenv}{Definition}
\newtheorem{lemmaenv}[definitionenv]{Lemma}
\newtheorem{theoremenv}[definitionenv]{Theorem}
\newtheorem{corollaryenv}[definitionenv]{Corollary}
\newtheorem{propositionenv}[definitionenv]{Proposition}
\newtheorem{conjectureenv}[definitionenv]{Conjecture}
\newtheorem{remarkenv}[definitionenv]{Remark}
\newenvironment{remark}{\begin{remarkenv}\rm}{\end{remarkenv}}
\newcommand{\br}{\begin{remark}}
\newcommand{\er}{\end{remark}}

\newtheorem{exampleenv}{Example}
\newtheorem{app-lemmaenv}[section]{Lemma}

\newenvironment{definition}{\begin{definitionenv}\rm}{\end{definitionenv}}
\newenvironment{lemma}{\begin{lemmaenv}\rm}{\end{lemmaenv}}
\newenvironment{theorem}{\begin{theoremenv}\rm}{\end{theoremenv}}
\newenvironment{corollary}{\begin{corollaryenv}\rm}{\end{corollaryenv}}
\newenvironment{example}{\begin{exampleenv}\rm}{\end{exampleenv}}
\newenvironment{proposition}{\begin{propositionenv}\rm}{\end{propositionenv}}
\newenvironment{conjecture}{\begin{conjectureenv}\rm}{\end{conjectureenv}}
\newenvironment{app-lemma}{\begin{app-lemmaenv}\rm}{\end{app-lemmaenv}}

\newcommand{\bd}{\begin{definition}}
\newcommand{\ed}{\end{definition}}
\newcommand{\bl}{\begin{lemma}}
\newcommand{\el}{\end{lemma}}
\newcommand{\elp}{\hspace*{\fill} $\Box$
                 \end{lemma}}
\newcommand{\bt}{\begin{theorem}}
\newcommand{\et}{\end{theorem}}
\newcommand{\etp}{\hspace*{\fill} $\Box$
                 \end{theorem}}
\newcommand{\bc}{\begin{corollary}}
\newcommand{\ec}{\end{corollary}}
\newcommand{\ecp}{\hspace*{\fill} $\Box$
                 \end{corollary}}
\newcommand{\bcj}{\begin{conjecture}}
\newcommand{\ecj}{\end{conjecture}}

\newcommand{\be}{\begin{example}}
\newcommand{\ee}{\end{example}}
\newcommand{\eep}{\hspace*{\fill} $\Box$
                 \end{example}}
\newcommand{\bp}{\begin{proposition}}
\newcommand{\ep}{\end{proposition}}
\newcommand{\epp}{
                 \end{proposition}}

\newcommand{\bra}[1]{\langle#1|}
\newcommand{\ket}[1]{|#1\rangle}

\newcommand{\tr}[1]{\text{tr}\left(#1\right)}

\newcommand{\eeq}{ \setcounter{equation} {\value{enumi}}}

\newcommand{\cA}{\mathcal{A}}
\newcommand{\cB}{\mathcal{B}}
\newcommand{\cC}{\mathcal{C}}

\newcommand{\cE}{\mathcal{E}}

\newcommand{\cG}{\mathcal{G}}
\newcommand{\cH}{\mathcal{H}}

\newcommand{\cM}{\mathcal{M}}

\newcommand{\cP}{\mathcal{P}}

\newcommand{\cR}{\mathcal{R}}

\newcommand{\sfF}{\textsf{F}}

\newcommand{\sfH}{\textsf{H}}

\newcommand{\mC}{{\mathbb C}}

\newcommand{\mR}{{\mathbb R}}

\def\tr{\textnormal{tr}}
\def\beq{\begin{equation}}
\def\eeq{\end{equation}}

\def\bean{\begin{IEEEeqnarray*}{rCl}}
\def\eean{\end{IEEEeqnarray*}}

\newcommand{\kg}{\textsf{KeyGen}}
\newcommand{\eval}{\textsf{Eval}}
\newcommand{\dec}{\textsf{Dec}}
\newcommand{\enc}{\textsf{Enc}}

\newcommand{\cnot}{\mathrm{CNOT}}

\newcommand{\triv}{\textsf{F}}
\newcommand{\etal}{{\it et~al.}}

\newcommand{\iqp}{\text{IQP}}

\newcommand{\itbms}{\text{IT $N_\kappa$-BMS}}

\ifnum\final=0
\newcommand{\mynote}[2]{{\color{#1} \marginpar{\tiny #2}}}
\newcommand{\mybignote}[2]{{\color{#1} $\langle \langle$ #2$\rangle \rangle$}}
\newcommandx{\rednote}[2][1=]{\todo[linecolor=red,backgroundcolor=red!25,bordercolor=red,#1]{#2}}
\newcommandx{\bluenote}[2][1=]{\todo[linecolor=blue,backgroundcolor=blue!25,bordercolor=blue,#1]{#2}}
\newcommandx{\yellownote}[2][1=]{\todo[linecolor=yellow,backgroundcolor=yellow!25,bordercolor=yellow,#1]{#2}}
\newcommandx{\greennote}[2][1=]{\todo[inline,linecolor=olive,backgroundcolor=green!25,bordercolor=olive,#1]{#2}}

\else
\newcommand{\mynote}[2]{}
\newcommand{\mybignote}[2]{}
\newcommand{\rednote}[2][1=]{}
\newcommand{\bluenote}[2][1=]{}
\newcommand{\greennote}[2][1=]{}
\newcommand{\yellownote}[2][1=]{}

\fi

\begin{document}
\title{On Statistically-Secure  Quantum  Homomorphic Encryption}

\author{Ching-Yi Lai}
\email{cylai0616@gmail.com}
\affiliation{\footnotesize Institute of Communications Engineering, National Chiao Tung University, Hsinchu 30010, Taiwan}
\author{Kai-Min Chung}
\affiliation{\footnotesize Institute of Information Science, Academia Sinica, Taipei 11529, Taiwan}


\begin{abstract}
Homomorphic encryption  is an encryption scheme that allows computations to be evaluated on encrypted inputs without knowledge of their raw messages.
Recently
Ouyang \emph{et al.} constructed a quantum homomorphic encryption  (QHE) scheme for Clifford circuits with statistical security (or information-theoretic security (IT-security)).
It is desired to see whether an information-theoretically-secure (ITS) quantum FHE exists. If not, what other nontrivial class of quantum circuits can be homomorphically evaluated with IT-security?
We provide a limitation for the first question  that an ITS quantum FHE necessarily incurs exponential overhead.
As for the second one, we propose a QHE scheme for the  instantaneous quantum polynomial-time (IQP) circuits.
Our QHE scheme for IQP circuits follows   from the one-time pad.
\end{abstract}

\maketitle
\vspace{-1cm}
\section{Introduction}
Homomorphic encryption  is an encryption scheme that allows computation to be evaluated on encrypted inputs without knowledge of their raw messages.
An encryption scheme  is typically considered as a computational primitive, since an information-theoretically-secure (ITS for short) symmetric key encryption scheme can only securely encrypt messages of length at most the length of the secret key  by Shannon's impossibility result~\cite{Shan49,Dodis2012}.
However, homomorphic encryptions can be interesting even with a bounded number of encrypted messages.
Unfortunately, classical ITS \emph{fully} homomorphic encryptions do not exist~\cite{KO97}.

In this work we investigate the possibility of  ITS  symmetric-key homomorphic encryptions in the quantum setting.
The quantum analogue of Shannon's impossibility result says that
no quantum encryption scheme with information-theoretic security (IT-security) can encrypt a message much longer than the secret key~\cite{AMTW00,DD10,LC18}.
In light of the negative result, we consider ITS  symmetric-key quantum homomorphic encryption (QHE) with  bounded-message security.
Such a QHE scheme is called IT $N_\kappa$ bounded-message secure (\itbms)  for a security parameter $\kappa$ (see Def.~\ref{def:it-secure}).
We remark that  security with respect to chosen-plaintext attacks (CPAs) where an adversary has access to an encryption oracle  is usually considered in the computational setting~\cite{BJ15,DSS16}; however, CPA security cannot hold in the information-theoretic setting  due to Shannon's impossibility results.

If a QHE scheme supports homomorphic evaluation of arbitrary quantum computation, it is called a quantum \emph{fully} homomorphic encryption (QFHE) scheme. For this, homomorphic evaluation of a universal set of gates need to be implemented efficiently. It is known that \emph{Clifford} gates together with the $T$ gate are universal for quantum computation.
The Clifford gates are composed of Hadamard,
phase,
and controlled-NOT $\cnot=\ket{0}\bra{0}\otimes I+ \ket{1}\bra{1}\otimes X$.
(Pauli matrices are denoted by $I,$ $X$, $Y$, and $Z$.)

In the computational setting, there has been an exciting development of QHE, initiated by the seminal work of Broadbent and Jeffery~\cite{BJ15}, who constructed (computationally secure) QHE schemes for Clifford circuits with a small number of $T$ gates. The first (leveled) QFHE is constructed by Dulek, Schaffner, and Speelman~\cite{DSS16} based on the learning with error (LWE) assumption. However, their scheme requires a large quantum evaluation key that is consumed in the homomorphic evaluation.
Alagic \etal~\cite{ADSS17} showed how to further achieve verifiability for QFHE, which allows verification of correctness of  homomorphically evaluated ciphertexts.
Very recently, Mahadev~\cite{Mah17} constructed a new QFHE scheme that significantly improved the result of~\cite{DSS16} by removing the quantum evaluation key. Furthermore, the scheme of Mahadev~\cite{Mah17} produces classical ciphertexts when the underlying message is classical, which allows a classical client to delegate quantum computation to a quantum server without revealing his classical input.

A natural question is: can we build a  QFHE scheme with IT-security?
The possibility was first investigated by Ouyang \emph{et al.}~\cite{OTF15} and they showed that ITS QHE can be performed
for  Clifford circuits.
On the other hand, it has been shown by Yu \emph{et al.}~\cite{YPF14} that  QFHE with perfect security must incur  an exponential overhead.
Herein, we extend the result to QFHE with imperfect IT-security
by a reduction to the communication lower bound of  \emph{quantum private information retrieval} (QPIR)~\cite{CKGS98,BB15}.
(Please see Theorem~\ref{thm:impossibility}.)
This limitation is also independently observed by Newman and Shi~\cite{NS17}.
As a consequence,   ITS QFHE does not exist.

The next question   is whether we can have ITS QHE  for any  nontrivial class
of  circuits other than the Clifford circuits.
In \cite{OTF15} a class of Calderbank-Shor-Steane (CSS)  codes~\cite{CS96,Ste96} with transversal Clifford gates are used to construct an ITS QHE scheme.
Quantum   codes are used in the setting of fault-tolerant quantum computation~\cite{DS96}
and quantum cryptography ({e.g.}, \cite{SP00,BCGST02,BCGHS06,BFK09,BGS13}).
Therefore, if we replace those CSS codes by codes with a different transversal gate set,
we obtain an ITS QHE scheme for another class of circuits.
For example, the triorthogonal codes~\cite{BH12} have transversal $\cnot$, $T$, and control-control-phase gates.
(However, it is known that  transversal gates alone cannot be universal for quantum   codes~\cite{EK09,ZCC11}.
See more discussion about transversal computation in~\cite{NS17}.)

In this paper
 we show how to do QHE for
 the  instantaneous quantum polynomial-time (IQP) circuits~\cite{SB09}.
{Note that the encrypted messages to IQP are restricted to the those density operators without any single Pauli $Z$ in the Pauli decomposition. See Def.~\ref{def:iqp} and the discussions there.}
We will propose a QHE scheme $\triv$ for $\iqp$ based on $Z$ one-time pads.

The notion of IQP computation   was proposed in \cite{SB09}, which is not universal for quantum computation.
It is known that  the class of IQP with \emph{postselection} is equivalent to the class PP~\cite{BJS10}.
Moreover, IQP computations are difficult to simulate with classical computers \cite{BJS10,BMS16,BMS17}
  unless the polynomial hierarchy  collapses to the third level.
Currently it is hard to implement Shor's factoring algorithm \cite{Shor94}. Instead, nonuniversal circuits, such as IQP, are physically more feasible
so that quantum supremacy could be demonstrated \cite{QLM+16,BMS17,Fujii16,GWD17}.
In contrast, a Clifford circuit with input states in the computational basis can be classically simulated by the Gottesman-Knill theorem~\cite{GK98}.

This paper is organized as follows.
Preliminaries are given in the next section, including basics of quantum information processing.
In Sec.~\ref{sec:def} we define QHE and its properties and then provide the limitation of IT-secure QFHE in Sec.~\ref{sec:limitation}.
In Sec.~\ref{sec:qhe iqp} we discuss IQP circuits and propose  the  QHE scheme $\triv$.
Then we conclude.

\section{Preliminaries} \label{sec:prelim}

We give notation and briefly introduce basics of quantum mechanics here.
A quantum system will be denoted by a capital letter and its corresponding Hilbert space will be denoted by the corresponding calligraphic letter.

Let $L(\cH)$ denote the space of linear operators on a complex Hilbert space $\cH$. A  quantum system is described by a \emph{density operator} $\rho\in L(\cH)$ that is positive semidefinite  and with trace one $\tr{\rho}=1$.
Let $D(\cH)= \{ \rho\in L(\cH): \rho\geq 0, \tr{\rho}=1\}$ be the set of density operators on a $\cH$.
When $\rho\in D(\cH)$ is of rank one, it is called a \emph{pure} quantum state and we can write $\rho=\ket{\psi}\bra{\psi}$ for some unit vector $\ket{\psi}\in \cH$,
where $\bra{\psi}=\ket{\psi}^{\dag}$ is the conjugate transpose of $\ket{\psi}$. If  $\rho$ is not pure, it is called a \emph{mixed} state and can be expressed as a convex combination of pure quantum states.
The Hilbert space of a joint quantum system $AB$ is the tensor product of the corresponding Hilbert spaces $\cA\otimes \cB$.
We use $\rho_{MR}$ to denote a density operator for the joint of the message (M) and reference (R) systems.

The trace distance between two quantum states $\rho$ and $\sigma$ is
$$||{\rho}-{\sigma}||_{\mathrm{tr}},$$
where $||X||_{\mathrm{tr}}=\frac{1}{2}\tr{\sqrt{X^{\dag}X}}$ is the trace norm of an operator $X$.

Associated with an $m$-qubit quantum system is a complex Hilbert space $\mC^{2^m}$ with a computational basis $\{\ket{v}:v\in \{0,1\}^m\}$.
Let $\{\ket{0},\ket{1}\}$ be an ordered basis for pure single-qubit states in $\mC^2$.
The Pauli matrices
\begin{align*} &\sigma_0=I=\begin{bmatrix}1 &0\\0&1\end{bmatrix},\ \sigma_1=X=\begin{bmatrix}0 &1\\1&0\end{bmatrix},\  \sigma_3=Z=\begin{bmatrix}1 &0\\0&-1\end{bmatrix},\ \sigma_2=Y=iXZ
  \end{align*}
form a basis of  $L(\mC^2)$.
 Then any single-qubit density operator $\rho\in D(\mC^2)$ admits a Bloch sphere representation \begin{align} \rho= \frac{I+r_1 X+ r_2 Y +r_3 Z  }{2}\triangleq  \frac{I+\vec{r}\cdot \vec{\sigma}  }{2}, \end{align} where $\vec{\sigma}=(\sigma_1,\sigma_2,\sigma_3)$ and $\vec{r}=(r_1,r_2,r_3)\in \mR^3$ is called the Bloch vector of $\rho$ such that $r_1^2+r_2^2+r_3^2\leq 1$. If $\rho$ is pure, we have $r_1^2+r_2^2+r_3^2= 1$.

The evolution of a quantum state $\rho\in D(\cH)$ is described by a \emph{quantum operation} $\cE: D(\cH)\rightarrow D(\cH')$ for some Hilbert spaces $\cH$ and $\cH'$.
In particular, if the evolution is a unitary $U$, we have the evolved state $\cE(\rho)=U\rho U^{\dag}$.
A quantum operation of several single-qubit Pauli operators on $n$ different qubits simultaneously can be realized as an $n$-fold Pauli operator. Denote  the $n$-fold Pauli group by  $${\mathcal{G}}_n=\{i^c E_1\otimes  \cdots \otimes E_n: c\in\{0,1,2,3\}, E_j\in\{I,X,Y,Z\} \}.$$
All elements in ${\cG}_n$ are unitary with eigenvalues $\pm 1$ and they either commute or anticommute with each other.  An $n$-fold Pauli operator admits a binary representation that is irrelevant to its phase. For two binary $n$-tuples $u,v\in \{0,1\}^n$, define
 \[
 Z^{u}X^{v}= \bigotimes_{j=1}^n Z^{u_j}X^{v_j}.
 \]
where $u= u_1\cdots u_n$ and  $v= v_1\cdots v_n$.
Thus any  $g\in {\mathcal{G}}_n$ can be expressed as $g= i^c Z^{u}X^{v}$ for some $ c \in \{0,1,2,3 \}$ and $u,v\in \{0,1\}^n$.

 The set of unitary operators in $L(\mC^{2^n})$ that preserve the $n$-fold Pauli group $\cG_n$ by conjugation is
 the  \emph{Clifford group}, which is generated by the Hadamard $(H)$,  phase $(P)$  and controlled-NOT ($\cnot$) gates:
\begin{align*}
H&= \frac{1}{\sqrt{2}}\begin{bmatrix}1 &1\\1&-1\end{bmatrix},\
P=\begin{bmatrix}1 &0\\0& i\end{bmatrix}, \ \cnot=\ket{0}\bra{0}\otimes I+ \ket{1}\bra{1}\otimes X.
 \end{align*}
The gates $H$, $P$, and $\cnot$ are called \emph{Clifford  gates}.
It is known that circuits composed of only Clifford gates are not universal; the Clifford gates together with any gate outside the Clifford group will do.
For example,
a candidate is the $\pi/8$ gate
\begin{align*}
   T&=e^{i\pi/8}\begin{bmatrix}e^{-i\pi/8} &0\\0& e^{i\pi/8}\end{bmatrix}.
 \end{align*}
These gates that involve only a few qubits are called elementary gates.
Then a quantum circuit is composed of a sequence of elementary gates and possibly some quantum measurements.
It is known that quantum measurements can be deferred to the end of a quantum circuit \cite{NC00} and we will assume it is always the case in this paper.
Also we consider only measurements in the $Z$ basis $(\ket{0},\ket{1})$ and measurements in the $X$ basis $(\ket{+}=\frac{\ket{0}+\ket{1}}{\sqrt{2}}, \ket{-}=\frac{\ket{0}-\ket{1}}{\sqrt{2}})$.
We denote by $C(\rho)$ the output of a quantum circuit $C$ with input quantum state $\rho$ by treating $C$ as a quantum operation.

\section{Definitions}\label{sec:def}

\begin{definition} \label{def:FHE}
A private-key quantum homomorphic qubit-encryption  scheme $\sfF$ is defined by the following   algorithms:
\begin{enumerate}[1)]

\item  (Key generation) $\sfF.\kg$: $1^{\kappa}\rightarrow  \cH_{\text{sk}} \times D(\cH_{\text{evk}})$. The algorithm takes an input of a security parameter $\kappa$ and outputs a \emph{classical} private key $sk$
        (and possibly some quantum evaluation key $\rho_{\text{evk}}\in D(\cH_{\text{evk}})$).

\item  (Encryption) $\sfF.\enc_{sk}$: $D(\mC^2)\rightarrow D(\cC)$.
The algorithm takes $sk$ and a single-qubit   $\rho\in D(\mC^2)$ as input and outputs a ciphertext  $\tilde{\rho}\in D(\cC)$.

\item (Evaluation) $\sfF.\eval$:  $D(\cH_{\textrm{evk}})\times \mathfrak{C}_{\kappa}\times D(\cC^{\otimes M_{\kappa}})\rightarrow D(\cC^{\otimes M_{\kappa}})\times \cM^{\otimes M_{\kappa}}$,
  where $\mathfrak{C}_{\kappa}$ is a set of admissible quantum circuits for $\sfF$ and the security parameter $\kappa$,  $M_{\kappa}$ is the maximum number of input qubits for $\mathfrak{C}_{\kappa}$,
  and $\cM$ is the set of corresponding homomorphic measurement outcomes.

\item (Decryption) $\sfF.\dec_{sk}$: $D(\cC)\times \cM\rightarrow D(\mC^2) \times \{0,1\}$.
If the measurement outcome $\mu\in \cM$ is trivial,
 the algorithm takes $sk$ and $\sigma\in D(\cC)$ as input and outputs a single-qubit quantum state $\hat{\sigma}\in D(\mC^2)$.
 Otherwise,  the algorithm takes $sk$ and $\mu$ as input and outputs a bit of measurement outcome.

\item[-](Correctness)
 $\sfF$ is \emph{homomorphic} for $\mathfrak{C}=\cup_{\kappa} \mathfrak{C}_{\kappa}$ if  there exists
a negligible function $\eta(\kappa)$ such that for $sk,\rho_{\text{evk}}\leftarrow \sfF.\kg(1^{\kappa})$,  $C\in \mathfrak{C}_{\kappa}$ on $M_{\kappa}$ qubits, and
$\rho_{MR}\in D(\mC^{2^{M_\kappa}}\otimes \cR)$,
\begin{align}
&\left\| { \Phi_{\sfF.\dec}^{sk, M_\kappa} \otimes \mathbb{I}_R \left(   \Phi_{\sfF.\eval}^{\rho_{\text{evk}},{C}} \otimes \mathbb{I}_R \left( \Phi_{\sfF.\enc}^{sk, M_\kappa}\otimes \mathbb{I}_R(\rho_{MR})\right)\right)} - { C\otimes \mathbb{I}_R ( \rho_{MR})} \right\|_{\mathrm{tr}}\notag \leq  \eta(\kappa),
\end{align}
where $\Phi$ denotes the corresponding quantum operation of the underlying algorithm and $R$ is a reference system.

\item[-] (Compactness) $\sfF$ is \emph{compact} for $\mathfrak{C}=\cup_{\kappa} \mathfrak{C}_{\kappa}$ if there exists a polynomial $p$ in $\kappa$ such that for any $C\in\mathfrak{C}_{\kappa}$, the circuit complexity of applying $\sfF.\dec$ to the output of $\sfF.\eval_{C}$ is at most $p(\kappa)$.
\end{enumerate}
\end{definition}

  A QHE scheme   is \emph{fully homomorphic} if it is compact and homomorphic for all quantum circuits generated by a universal set of quantum gates.\\

\begin{definition}\label{def:it-secure-s}
  A QHE scheme $\sfF$ is \emph{information-theoretically $N_{\kappa}$-bounded-message-secure} (\itbms)   if   there exists a negligible function $\epsilon(\kappa)$,
such that for every security parameter $\kappa$  and $\rho_{MR},\rho'_{MR}\in D(\mC^{2^{N_\kappa}}\otimes \cR)$ with $\tr_M(\rho_{MR})=\tr_M(\rho'_{MR})$,
  \begin{align}
 \left\|     \Phi_{\sfF.\enc}^{sk, N_\kappa} \otimes \mathbb{I}_R (\rho_{MR}) -    \Phi_{\sfF.\enc}^{sk, N_\kappa} \otimes \mathbb{I}_R(\rho'_{MR}) \right\|_{\mathrm{tr}}
&\leq  \epsilon(\kappa),
\end{align}
where $\Phi$ denotes the corresponding quantum operation of the underlying algorithm.
\end{definition}

\noindent
This definition says that two encrypted quantum states of an IT-secure QHE are statistically indistinguishable.

Below is a weaker definition of $\itbms$ with no reference system involved.
\begin{definition}\label{def:it-secure}
  A QHE scheme $\sfF$ is \emph{weak} \emph{information-theoretically $N_{\kappa}$-bounded-message-secure} (weak \itbms)   if   there exists a negligible function $\epsilon(\kappa)$,
such that for every security parameter $\kappa$ and $\rho,\rho'\in D(\mC^{2^{N_\kappa}})$,  
  \begin{align}
\left\|   \Phi_{\sfF.\enc}^{sk, N_\kappa}(\rho)-   \Phi_{\sfF.\enc}^{sk, N_\kappa}(\rho')\right\|_{\mathrm{tr}}\leq \epsilon(\kappa). \label{eq:4}
  \end{align}
\end{definition}

It is clear that Def.~\ref{def:it-secure-s} implies  Def.~\ref{def:it-secure} by choosing a trivial reference system.
By \cite[Theorem 8]{LC18}, we have the following theorem.
\bt \label{thm:weak to strong}
If $\sfF$ is a weak $\itbms$ encryption scheme with security error at most $\epsilon(\kappa)$, then $\sfF$ is $\itbms$ with   security error at most $2^{2N_\kappa+1} \epsilon(\kappa)$.
\et

\section{Limitation}\label{sec:limitation}

The negative result for ITS QFHE is proved by a reduction to the communication lower bound of quantum private information retrieval (QPIR) with one server.
In an $(n,m)$  QPIR problem, Alice (the server) has a data string $x$ of $n$ bits  and Bob wishes to learn the $i$th entry $x_i$.
They create an initial state $\rho_{ABR}$ with $\tr_R \rho_{ABR}=\ket{x}_A\bra{x}\otimes \ket{i}_B\bra{i}$ and after a two-party protocol $\cP$ they end up with the final state $\rho'_{ABR}=\cP(\rho_{ABR})$
by exchanging $m$ qubits.
We say that $\cP$ has \emph{correctness error} $\delta$ if for any $\rho_{ABR}$ with $\tr_R \rho_{ABR}=\ket{x}_A\bra{x}\otimes \ket{i}_B\bra{i}$, there exists a measurement $\cM$ such that
\begin{align*}
& \Pr\left\{ \cM\left( \tr_{AR}(\rho'_{ARB}) \right)=x_i \right\}\geq 1-\delta.
\end{align*}
We say that $\cP$ has \emph{security error} $\epsilon$ if for any $\rho_{ABR}$, there exists a quantum operation  $\cE_{AR}$ such that
\begin{align*}
&\left\|    \tr_B \left(\cE_{AR} \otimes \mathbb{I}_B (\rho_{ARB})\right) -    \tr_{B}(\rho'_{ARB}) \right\|_{\mathrm{tr}}\leq \epsilon.
\end{align*}
Nayak~\cite{Nayak99,ANTV02} proved  that $m\geq (1-H(1-\delta))n$ for $\epsilon=0$, where $H(p)$ is the binary entropy function. This was extended to the case of $\epsilon>0$ in \cite{BB15}:
\begin{align}
m\geq \left(1-H\left(1-\delta-2\sqrt{\epsilon(1-\epsilon)}\right)\right)n. \label{eq:nayak_bound}
\end{align}
We show that the existence of an  ITS QFHE would contradict this communication lower bound.
\bt \label{thm:impossibility}
There is no  $\itbms$ QFHE for $N_{\kappa}=\omega(\log \kappa)$ with correctness error $\eta(\kappa)=0.0001$ and security error $\epsilon(\kappa)=0.0001$.
Moreover, there is no  $\itbms$ QHE on classical circuits for $N_{\kappa}=\omega(\log \kappa)$ with correctness error $\eta(\kappa)=0.0001$ and security error $\epsilon(\kappa)=0.0001$.

\et
\begin{proof}

Assume there is an $\itbms$ QFHE scheme $\sfF$ for $N_{\kappa}=\omega(\log \kappa)$
with correctness error $\eta(\kappa)=0.0001$ in Def.~\ref{def:FHE} and security error $\epsilon(\kappa)=0.0001$ in Def.~\ref{def:it-secure}.
 We show that it leads to a QPIR protocol that violates  the communication lower bound Eq.~(\ref{eq:nayak_bound}).

Suppose that $\sfF$ encrypts one qubit into a ciphertext of at most $p(\kappa)$ qubits and $\kappa$ is  sufficiently large such that $$H\left(1-\eta(\kappa)-2\sqrt{\epsilon(\kappa)(1-\epsilon(\kappa))}\right)<0.2.$$
Let $n=100 p^2(\kappa)$, which satisfies $\log n\leq N_\kappa$.

Suppose Alice holds a database $x\in\{0,1\}^n$ and Bob wants to retrieve information $x_i$ from Alice by using $\sfF$ without revealing   $i$.
 Let $C_x\in \mathfrak{C}_{\kappa}$ be a quantum circuit that takes an input $i\in\{1,2,\dots,n\}$ and outputs  $x_i$.

The QPIR protocol is as follows.
 Suppose Bob wants to query an index $i^*$ of $\log n$ bits.  Using the QFHE algorithm $\sfF$,  he generates a   key set $sk, \rho_{\text{evk}}\leftarrow \sfF.\kg(1^{\kappa})$
and then produces the cipher state
$\sfF.\enc_{sk}^{\otimes \log n}(\ket{i^*}),$
 which is of $p(\kappa)  \log n$
qubits. Then  he sends it to Alice, together with $ \rho_{\text{evk}}$.
After computing  $$\rho \triangleq \sfF.\eval(\rho_{\text{evk}},{C_x},\sfF.\enc_{sk}^{\otimes \log n}(\ket{i^*})),$$
Alice sends $\rho$ back to Bob.
By the security of $\itbms$ $\sfF$, Bob does not reveal his desired objective to Alice.
If Alice honestly does the homomorphic evaluation, Bob would learn $x_{i^*}$ with probability at least $1-\eta(\kappa)$. Thus we have an $\left(n=100p^2(\kappa), p(\kappa)(1+\log n)\right)$ QPIR protocol with correctness error $\eta(\kappa)$ and security error $\epsilon(\kappa)$.
Note that only  $p(\kappa)(1+\log n)=p(\kappa)(1+\log 10 +2\log p(\kappa))$ qubits are required in communication,
which contradicts the communication lower bound~(\ref{eq:nayak_bound}) that at least $0.8 n=80p^2(\kappa)$ qubits are required.

Observe that in this proof Alice only needs to homomorphically evaluate  a classical selection function. Thus the impossibility proof also rules our ITS QHE schemes for all classical circuits.
\end{proof}

\section{ IQP  Circuits}\label{sec:qhe iqp}
IQP circuits are proposed in \cite{SB09,BJS10}.
We call a gate diagonal in the computational basis $(\ket{0},\ket{1})$  a \emph{diagonal gate}.
\begin{definition} \label{def:iqp}
An IQP circuit  is a quantum circuit consisting of  diagonal gates.
The input state is the product state of some qubits in $\ket{+}$ and
the output is the measurement outcomes on a specified subset of the qubits in the $X$ basis  $(\ket{+},\ket{-})$.
\end{definition}

\br
 For the homomorphic encryption of $N$-qubit IQP   circuits, we define a more general input  space:
\begin{align}
D_{xy}(\mC^{2^N}\otimes \cR)\triangleq \{   \rho_{MR}\in D(\mC^{2^N}\otimes \cR):  \rho_{MR}= \sum_i \alpha_{i} A_{i,1}\otimes \cdots\otimes A_{i,N} \otimes B_i, \mbox{ } A_{i,j}\in\{I,X,Y\}, B_i\in  D(\cR) \},
\end{align}
where $\cR$ is a reference system. \er
In other words, $D_{xy}(\mC^{2^N}\otimes \cR)$ is a collection of density operators that do not have any single $Z$ operator in the message part of their Pauli decompositions.
Similarly to Theorem~\ref{thm:weak to strong}, we have the following corollary.

\bc \label{cor:weak to strong}
If $\sfF$ is a weak $\itbms$ encryption scheme for input space $D_{xy}(\mathbb{C}^{N_\kappa})$ with security error at most $\epsilon(\kappa)$, then $\sfF$ is $\itbms$ for input space $D_{xy}(\mathbb{C}^{N_\kappa}\otimes \mathcal{R})$ with   security error at most $2^{2N_\kappa+1} \epsilon(\kappa)$.
\ec

 Then it suffices to prove that a QHE scheme for IQP  is weak $\itbms$. 

\subsection{QHE for IQP}
Observe that an input qubit of a IQP circuit lies in the $xy$-plane, i.e., $\ket{\psi}=\ket{0}+e^{i\theta}\ket{1}$,
which can be protected by a $Z$ one-time pad.
The symmetric-key QHE scheme  $\triv$ is  defined as follows.\\
$\overline{\hspace{\textwidth}}$\\
QHE scheme for IQP circuits: $\triv$\\
$\overline{\hspace{\textwidth}}$
Let $\text{IQP}_{N}$ be the set of IQP circuits with at most $N$ input qubits.
Suppose a client asks a server to compute a quantum circuit $C\in\text{IQP}_{N}$ on an  $N$-qubit input state in $ D_{xy}(\mathbb{C}^{2^ N})$.

\begin{enumerate}[1)]
\item $\triv.\kg : 1^{\kappa}\rightarrow h:\{0,1\}^\kappa\rightarrow\{0,1\}$, where  $h$ is
uniformly drawn from a family of  $\kappa$-independent  hash functions $\sfH_\kappa$ (no evaluation key here).

\item
$\triv.\enc$:  $\sfH_\kappa\times D(\mathbb{C}^2)\rightarrow \{0,1\}^\kappa \times D(\mathbb{C}^2)$. Encryption is done by applying a  $Z$ one-time pad:
 $$\triv.\enc_{h}(\rho)= \left(r, Z^{h(r)} \rho Z^{h(r)}\right)$$ for $\rho\in D(\mathbb{C}^2)$,
 where $r$ is a random string of $\kappa$ bits (encryption randomness).
For an $N$-qubit input state, each qubit is encrypted respectively.

\item $\triv.\eval$:  $\text{IQP}_{N}\times  \{0,1\}^{N\times \kappa} \times  D(\mC^{2^{ N}})\rightarrow \{0,1\}^{N\times \kappa} \times D(\mC^{2^{ N}})\times \{\bot,0,1\}^{N}$.
Evaluation is trivial for diagonal gates since they commute with the encryption.
Measurements in the $X$ basis are also trivial and the outcomes are protected by the  $Z$ one-time pads.
(Note that an additional register, initialized in $\bot$, is used the store the measurement outcomes $\{0,1\}$ for each qubit.)

 \item  $\triv.\dec_{h}$: $\{0,1\}^\kappa\times D(\mathbb{C}^{2})\times \{\bot,0,1\}\rightarrow D(\mC^{2})\times \{\bot,0,1\}$.
 For $r\in\{0,1\}^\kappa$, $\sigma\in D(\mathbb{C}^2)$, and $\mu\in\{\bot, 0,1\}$,
 if $\mu=\bot$, the qubit was not measured and the qubit state will be decrypted:
$\triv.\dec_h\left( (r, \sigma, \mu) \right)= Z^{h(r)} \sigma Z^{h(r)}$;
otherwise, the qubit is measured and the measurement outcome will be decrypted:
$\triv.\dec_h\left( (r, \sigma, \mu) \right)= h(r)\oplus \mu $.
\end{enumerate}
\noindent$\overline{\hspace{\textwidth}}$\\

Note that a $\kappa$-independent hash function $ h:\{0,1\}^\kappa\rightarrow\{0,1\}$ can be described in $O(N_\kappa+\kappa)$ bits (see, eg.,~\cite{Vad11}).
The secret keys used in $\triv$ are $k$-wise independent hash functions   with $k=O(N_\kappa)$
and the length of a secret key is $O(N_\kappa+\kappa)$.
The  fresh ciphertext has $N_\kappa$ qubits and $O\left( \kappa\right)$ classical bits.

\section{Discussion}

In this paper we proved a more general limitation on QFHE with imperfect IT-security,
which says that an exponential overhead is necessary even a small security error is allowed.
Our proof is based on a reduction to the communication lower bound of  quantum private information retrieval.

On the positive side, we provided a QHE scheme for IQP circuits based on Z one-time pads and hash functions.
The next question is whether we can allow verification of QHE for IQP circuits as in~\cite{ADSS17}.
We remark that it is also possible to derive a QHE for IQP based on quantum error-correction codes similar to the Clifford scheme by Ouyang \etal~\cite{OTF15}.
However, it is not clear how to introduce verifiability in the scheme.
On the other hand, the Clifford scheme by Ouyang~\etal naturally allows circuit privacy.
It is also unknown whether we can have circuit privacy for QHE on IQP circuits.

\section*{Acknowledgements}
We thank Si-Hui Tan for introducing their work~\cite{OTF15} to us. We are grateful to anonymous referees for their comments and the suggestion of the trivial scheme.
CYL would like to thank Todd\,B.\,Brun, Nai-Hui\,Chia, and Wei-Kai\,Lin for helpful discussions.
KMC acknowledges helpful discussions with Salil\,P.\,Vadhan.

CYL was was financially supported from the Young Scholar Fellowship Program by Ministry of Science and Technology (MOST) in Taiwan, under
Grant MOST107-2636-E-009-005.
KMC was partially supported by 2016 Academia Sinica Career Development Award under Grant
No. 23-17 and the Ministry of Science and Technology, Taiwan under Grant No. MOST 103-2221-
E-001-022-MY3.


\end{document}